# Device Improvement and Circuit Performance Evaluation of complete SiGe Double Gate Tunnel FETs

Bahniman Ghosh[1], Rahul Mishra[1]

[1]Department of Electrical Engineering , Indian Institute of Technology, Kanpur. Kanpur, Uttar Pradesh, India. (208016)
email : ramis@iitk.ac.in

**Abstract** – In recent part extensive simulation work has already been done on TFETs. However this is limited to device performance analysis. Evaluation of circuit performance is a topic that is very little touched. This is due to the non availability of compact models of Tunnel FETs in the commercial simulator. In our paper for the first time we perform the circuit analysis of tunnel FETs (extended channel TFETs), we test them over basic digital circuit. We generate the TFET models by using the model editor in Orcad. Extensive circuit simulation is then performed by using these models in the Pspice circuit design. Performance of extended channel double gate TFET is evaluated on the grounds of power and delay in inverter, nand gate, nor gate and ring oscillator. Before that we perform device analysis of double gate extended channel TFETs, extended channel has been tried before on SOI TFETs we try it for the first time on double gate $Si_{1-x}Ge_x$ TFETs. We even look at the effect of introducing Si layer. The performance of this device is compared for different Ge mole fraction and also with MOSFETs

## I. Introduction

Study on tunnel FETs have proved them to be better than conventional MOSFETs in terms of steeper sub threshold swing, higher $I_{on}/I_{off}$ ratio, lower power consumption (dynamic and leakage) and their scaling is not limited by the quantum mechanical effect unlike the MOSFETs [1]-[10]. Tunnel FETs are basically low power device. They work on the principle of band to band tunneling, from the valence band in source side to the conduction band in channel side [1]. The tunneling is initiated with the application of gate voltage, which lowers the tunneling distance between the two bands. Use of SiGe layer in the source side of tunnel FETs to increase the on current has already been proposed [9]-[12]. Similarly use of high K dielectrics like $HfO_2$ (K=22) and double gates have also helped in increasing the on current of TFETs [1]-[3]. Increasing Ge mole fraction in SiGe layer leads to higher on current by reducing the tunneling barrier due to lower band gap of SiGe [9]. An alternate to this is to use SiGe throughout the body. However with increasing Ge content the $I_{off}$ of such TFETs also increases proportionally which counteracts its advantage. A way to tackle this problem was proposed in [13] and was very nicely demonstrated by experimental results. However it dealt with SOIs and SiGeOIs. In our paper for the first time we provide a full insight on the use of this technique (extended channel) in double gate $Si_{1-x}Ge_x$ tunnel FETs and then perform circuit simulation of these TFETs. Double gate has the advantage of increased $I_{on}$ over SOI. We also optimize length of the extended channel and the Si layer which was used to further improve the performance. We compare extended tunnel FETs with different Ge mole fraction on the grounds of $I_{on}$, $I_{off}$, sub threshold swing etc. We also compare the performance of these TFETs with MOSFETs of similar dimension and identical threshold voltage. In later section we model tunnel FETs and use them in some basic digital circuits. We compare performance of these tunnel FETs for various Ge mole fraction and also compare them with corresponding MOSFETs.

## II. Device structure and simulation

A tunnel FET is basically a P-i-N junction device. With the n side serving as drain, p side as source and channel in intrinsic region in case of n-channel TFETs [1]. Gates are there on top of intrinsic region with a dielectric in between. In case of extended channel tunnel FET the gate does not cover the complete channel region. Fig. 1 shows the extended channel tunnel FET. The simulated tunnel FET is a complete SiGe tunnel FET. The substrate is P doped with carrier concentration of $10^{16}$ cm$^{-3}$. Gaussian profile is used in the source and drain region. The source region is P doped with peak density of

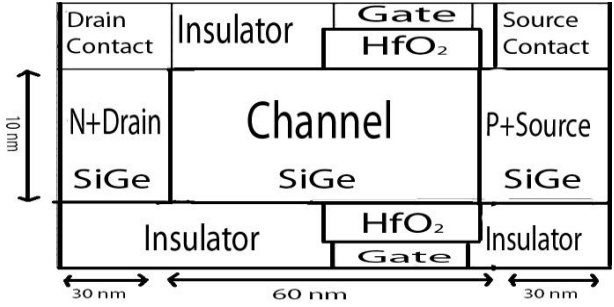

Fig.1. Extended channel SiGe N-tunnel FET

$10^{20}$ cm$^{-3}$ and with characteristic length of 0.55 nm. The drain region is N$^+$ doped with peak density of $10^{18}$ cm$^{-3}$ and characteristic length of 0.55 nm. For a P-TFET n-region is heavily doped compared to p-region. Optimum device parameters used in all the subsequent simulations are given in table 1. Metallic gate with workfunction 4.3 ev is used in case of N channel tunnel FET. All simulations were carried out using medici version Y-2006.06 A band to band tunneling model was used to account for the tunneling. In all the semiconductor regions where the current continuity equations are solved band-to-band tunneling generation is computed [14]. The drift diffusion model for current transport is used. A PTFET also has similar structure with N-type being the source and P-type as drain.

**Table I.** Device Parameter of NTFET used in simulation

| Ge mole fraction | 0 to 1 |
|---|---|
| Peak Source Doping (atoms/cm$^3$) | $10^{20}$ |
| Peak Drain Doping (atoms/cm$^3$) | $10^{18}$ |
| Substrate Doping (atoms/cm$^3$) | $10^{16}$ |
| Channel Length | 60nm |
| Dielectric thickness | 3 nm |
| Body thickness | 10 nm |
| Gate work function | 4.3 eV |

### III. Operation

As the name suggests, a tunnel FET works on the principle of electron tunneling. Fig. 2 shows the on and off state diagram of a simple TFET with Ge mole fraction of 0.8. As clear from the on state diagram, tunneling occurs at the source - channel junction as soon as the tunneling barrier reaches below required minimum. The tunneling current depends on the probability of electron tunneling from valence band in the source side to the conduction band in the channel. More the probability of tunneling more is the current. Among various other factors tunneling probability is dependent on the tunneling width (Fig. 2). Tunneling width will decrease as we decrease the band gap and hence it will increase the tunneling probability. Therefore SiGe with high Ge content has higher tunneling probability due to lower band gap. Equation (1) governs the tunneling current in a tunnel FET [15].

$$I \propto \exp\left[\frac{-4\sqrt{2m^*}E_g^{3/2}}{3e\hbar(\Delta\phi+E_g)}\sqrt{\frac{\varepsilon_{si}}{\varepsilon_{ox}}}\sqrt{t_{ox}t_{Si}}\right] \quad (1)$$

In the above equation $E_g$ is the band gap, $m^*$ is the effective carrier mass. $\Delta\phi$ is the energy range over which tunneling can take place. $t_{ox}$ and $t_{si}$ are dielectric and body thickness respectively. Rest of the symbols has their usual meanings. Thus it is clear from equation (1) that tunneling current can be changed by changing band gap. Effective masses are taken as $m_c=0.32m_o$ and $m_v=0.81m_o$ for Si$_{1-x}$Ge$_x$ with $x < 0.85$ and $m_c=0.22m_o$ and $m_v=0.34m_0$ for Si$_{1-x}$Ge$_x$ with $x > 0.85$ [16].

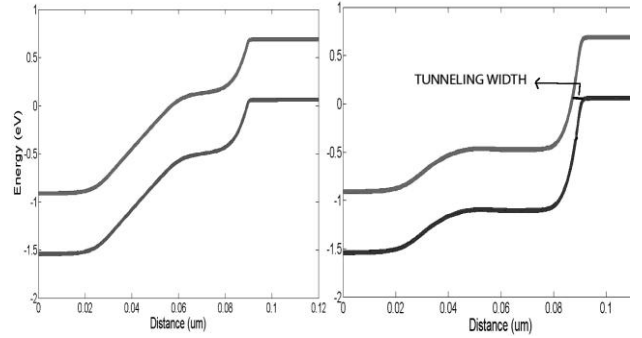

Fig. 2. Off state and On state band diagram for a SiGe extended TFET for Ge=0.8. Cross-section taken 2.5 nm from gate dielectric interface. For off state Vds= 1 volt . Vgs= 0 volt. For On state Vgs=Vds=1 volt.

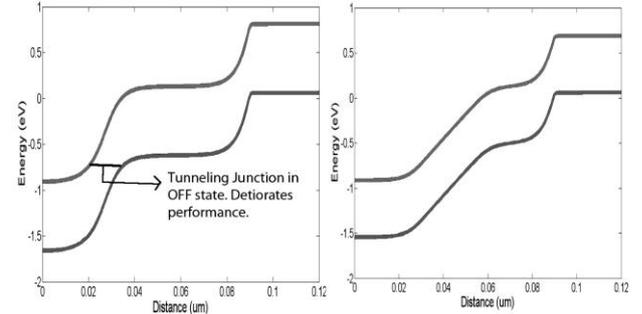

Fig. 3. Off state band diagram of normal (right) and extended channel tunnel FET (left). Cross-section taken 2.5 nm from the gate dielectric interface. For off state Vds= 1 volt . Vgs= 0 volt.

To compare between the extended channel TFET and the normal TFET we look at the off state diagram of both devices in Fig. 3 for Ge mole fraction of 0.8. In a normal tunnel FET as we increase the Ge mole fraction the tunneling barrier in the off state at the drain-channel junction decreases resulting in high off current[9]-[11]. From Fig. 3 we see that this drain-channel tunneling barrier is 0.01 μm for normal tunnel FET. However with the introduction of extended channel we are able to increase this tunneling barrier to 0.02 μm and hence we are able to reduce the electric field near the drain-channel junction [12]. This result in a lower OFF current compared to normal tunnel FET.

Before moving on to studying the characteristics of extended channel tunnel FET we look at the optimized channel length to get the desired performance. Because we are introducing the extended channel to improve the off current, we look at the variation of this off current

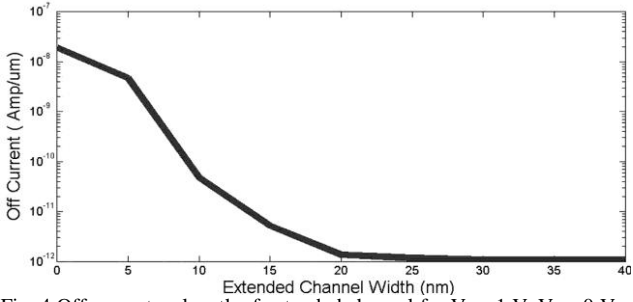

Fig. 4 Off current vs length of extended channel for $V_{ds}=1$ V, $V_{gs}=0$ V, Ge=0.8

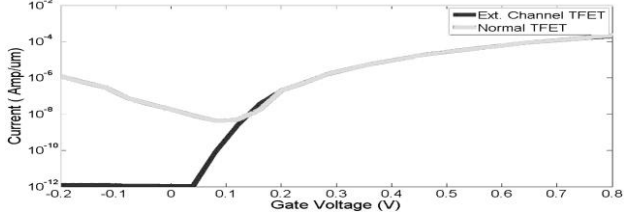

Fig. 5 Drain current vs Gate voltage for $V_{ds}=1$ V and Ge=0.8

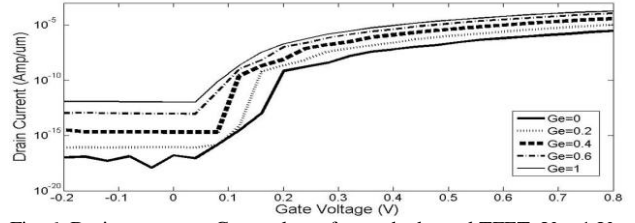

Fig. 6 Drain current vs Gate voltage for extd. channel TFET. $V_{ds}=1$ V.

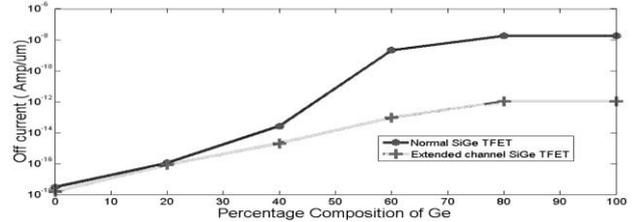

Fig. 7. $I_{off}$ vs Ge composition for $V_{ds}=1$ V. $V_{gs}=0$ V

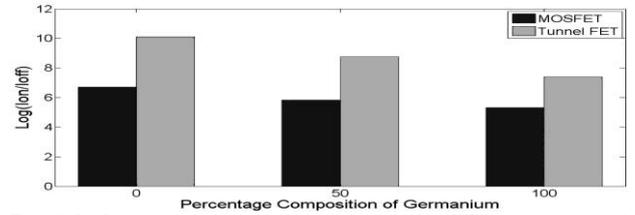

Fig. 8 $I_{on}$-$I_{off}$ ratio vs Ge composition for MOSFETS and extended channel TFETs. On state ($V_{ds}=1$ V, $V_{gs}=0.5$ V ). Off state($V_{gs}=0$)

with length of extended channel. Fig. 4 shows the variation. Clearly the off current saturated to a minimum for a extended channel length of 30nm. To keep the device dimension minimum we choose this width for subsequent simulations. In the next section we present the simulation result for optimized tunnel FET.

## IV. Simulation Results

Now we move on to comparing various characteristics of our extended channel TFET.

### A. Transfer characteristic

First we compare the $I_d - V_g$ characteristic of normal tunnel FET and extended channel TFET for Ge mole fraction of 0.8 for both case in Fig. 5. Fig. 6 compares the $I_d$-$V_g$ curves of extended channel tunnel FETs for various Ge mole fractions. Clearly increasing Ge mole fraction decreases the tunneling width due to decreasing band gap and hence increases the drain current. On current (at $V_{gs}=0.5$ V, $V_{ds}=1$ V) is $2.66\times10^{-5}$ A/μm for Ge=0.8 compared to $2.029\times10^{-7}$ A/μm for pure silicon. Increasing mole fraction of Ge also increases the off current. Off current is $1.09\times10^{-12}$ A/μm for Ge=0.8 which is much higher than the off current value of $1.9\times10^{-18}$ A/μm for pure silicon. Fig. 7 compares the off current of normal and extended channel tunnel FET for various Ge mole fractions. Cleary extended channel tunnel FETs have advantage over normal tunnel FETs at higher Ge mole fraction. Off current of normal TFET is $1.8\times10^{-8}$ A/μm for Ge=0.8 compared to $1.09\times10^{-12}$ A/μm for extended channel tunnel FET. As discussed earlier this is due to their larger tunneling width at drain-channel junction and hence reduced off current as clear from the figure. Fig. 8 depicts the $I_{on}$-$I_{off}$ ratio for various Ge mole fraction for extended channel tunnel FETs and similar size MOSFETs. TFETs have clear advantage over MOSFETs with their high $I_{on}$-$I_{off}$ ratio, this is due to their lower off current. We don't not compare the $I_{on}/I_{off}$ ratio of TFETs and MOSFETs for different Ge mole fraction. However we take a particular Ge mole fraction and then compare the performance of TFETs and corresponding MOSFETs. In short we are comparing performance of TFETs and MOSFETs for same material and not looking at their trend with change of material.

### A. Threshold Voltage

We define threshold voltage as the gate voltage at which drain current is equal to $10^{-7}$ A/um. As we see in Fig. 9 the threshold voltage decreases with increasing Ge mole fraction. This is due to the increasing current of the extended TFET with increasing Ge mole fraction. Threshold voltage is 0.18 V for extended TFET with Ge=0.8 compared to 0.46 V for pure silicon (gate work function = 4.3 eV). This threshold voltage is also dependent on the work function of the gate. Fig 10 shows the threshold variation with gate work function for an extended channel tunnel FET with Ge=0.8. As clear from the figure threshold voltage decreases with decreasing work function. Desired threshold voltage can be set for a given device by adjusting the workfuction of gate. However the subthreshold current must be kept in mind.

### B. Subthreshold swing

We now compare the performance of MOSFETs and TFETs in term of subthreshold swing for different germanium mole fraction. We however do not look at the variation of subthreshold swing with Ge mole fraction as MOSFETs and TFETs behave differently with varying Ge mole fraction. Performance of MOSFETs is limited by their 60mv/decade subthreshold swing. TFETs show a subthreshold swing below 60mV/dec thus making

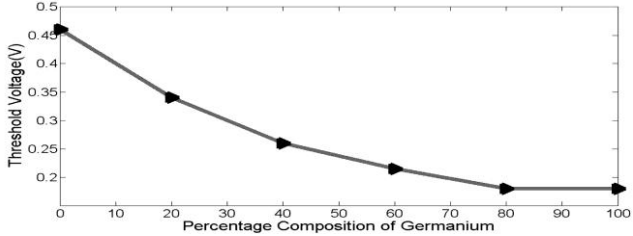
Fig. 9 Threshold voltage vs Ge composition for ext. channel TFETs.

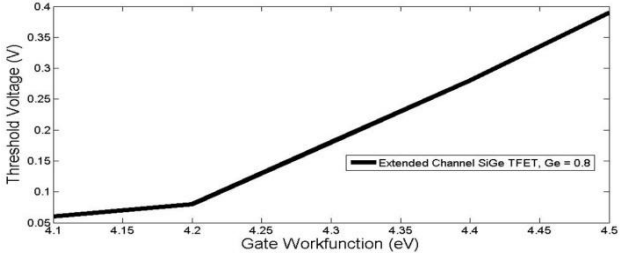
Fig.10. Threshold voltage vs Gate workfunction. Ge=0.8

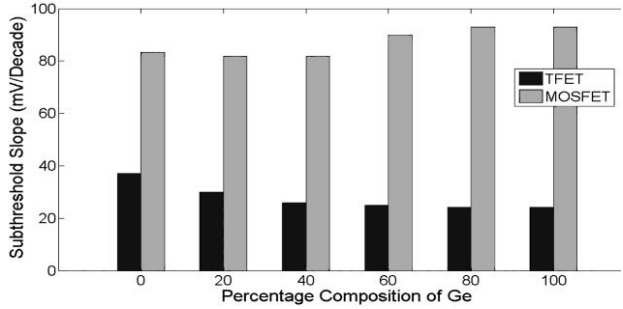
Fig.11 Average subthreshold slope vs Ge mole fraction.

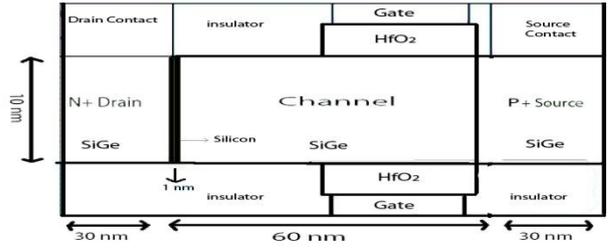
Fig. 12 Extended SiGe TFET with Si layer (1 nm)

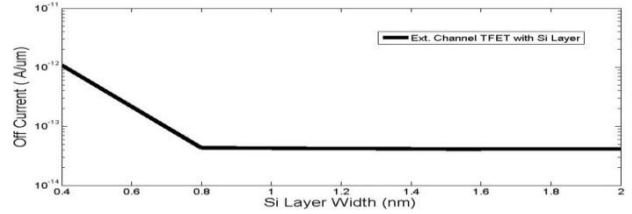
Fig. 13 Off current vs Si layer width for ext. channel TFET. Ge = 0.8

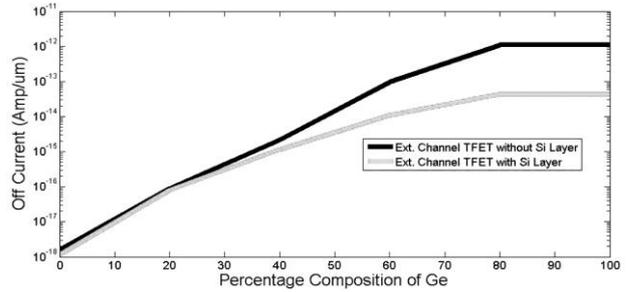
Fig.14 $I_{off}$ vs Ge mole fraction for Vds = 1 V. Vgs = 0 V

them a better switching device [1]. We take the definition of average subthreshold swing as given in equation (2) [15]. Here $V_t$ is threshold voltage, $V_{off}$ is the voltage at which TFET is off. $I_{vt}$ is the drain current at threshold ($10^{-7}$ A/μm in our case) and $I_{off}$ is the current in off state.

$$S = \frac{V_t - V_{off}}{\log I_{vt} - \log I_{off}} \quad (2)$$

Fig. 11 shows variation of subthreshold swing calculated using above formulae with Ge mole fraction for an extended channel TFET and a similar dimension MOSFET with their threshold voltage matched. Clearly TFETs show much lower subthreshold swing compared to MOSFETs (for Ge=0.8, subthreshold swing is 24 mV/Dec for TFET compared to 92 mV/Dec for MOSFET) and this subthreshold swing decreases with increasing Ge mole fraction but saturates after Ge=0.5 as the effect due to decreasing $V_t$ is compensated by increasing off current.

### C. Introduction of Si layer

Moving towards further decreasing the off current we introduce a layer of silicon at the drain-channel interface in a SiGe TFET with high Ge mole fraction. Fig. 12 shows the new structure. This is similar to what was done in SiGeOI in [13]. We try the same thing in a double gate extended channel tunnel FET. The combined effect of extended channel and Si layer further decreases the off current of SiGe TFET while keeping the on current same. The variation of $I_{off}$ with Si layer width is shown in Fig.13 for Ge mole fraction of 0.8. Care must be taken that Si layer does not extend till the source-channel junction as it will deteriorate the on current in that case. The off current saturates to a minimum for a Si width of 0.8 nm. We select Si width of 1 nm for our simulations. Fig. 14 compares the $I_{off}$ of the extended channel TFET with and without Si layer. For a Ge mole fraction of 0.8 TFET with Si layer has off current of $4.93 \times 10^{-14}$ A/μm compared to the $1.09 \times 10^{-12}$ A/μm of the one without Si.

## V. TFET modeling and circuit Simulation

In the further sections whenever we talk about TFETs we refer to the extended channel TFETs (without Si layer). We move towards analyzing the circuit performance of TFETs in terms of their power dissipation and delay. We use Cadence OrCAD V 16.0 [17]. However before that we must have appropriate models of TFETs so that their circuit can be simulated. Since the commercial circuit simulators don't provide us with compact model of TFET we have to generate it by ourselves. For this purpose we use the model editor available in OrCAD. Now, TFETs like MOFETs are three terminal device, namely the source, drain and gate. With the help of curve fitting capability of model editor we fit the current-voltage, transconductance and capacitance curves of a MOSFET to that of a TFET. The corresponding values for extended TFET were extracted using the device simulator (medici) and then fed in corresponding tables and then the curves were generated. The various curves that we got show good resemblance with the original device curves in the area of interest. Thus this MOSFET model with its curves fit to that of extended channel TFET was used for circuit simulation using OrCAD PSpice. From now on we call it as the TFET model.

Now the circuits in hand have both the P channel TFET and N channel TFET. The current voltage characteristic of these devices should be matched for best performance of circuits. Now a PTFET is similar to NTFET but with P side as drain and N side as source and N side heavily doped compared to P [6]. Unlike NTFET, the threshold voltage of PTFET decreases with increasing gate work function. Thus the threshold voltage and the doping level are adjusted to make the current – voltage characteristic of PTFET and NTFET symmetric. Fig. 15 shows one such match for SiGe P and NTFET with Ge mole fraction of 0.8.

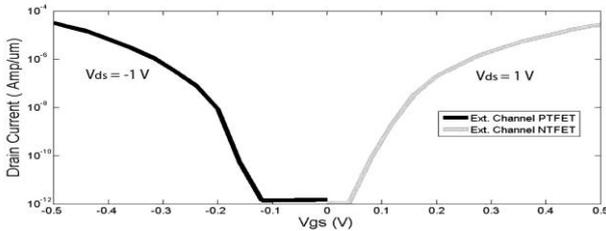

Fig. 15 Nearly symmetric I-V characteristic of P channel and N channel TFET. For PTFET, source doping ($N^+$) : $10^{20}$ cm$^{-3}$, drain doping ($P^+$): $10^{18}$ cm$^{-3}$. Gate work function 4.8 eV. Rest of the parameters same as in table 1. Parameters for NTFET same as in table 1.

Thus models of PTFETs and NTFETs were generated for various Ge mole fraction (0 , 0.2 , 0.4 , 0.6 , 0.8 and 1). And these models were then used for circuit simulation. The circuits that are analyzed are loaded inverter, nand gate and nor gate and 5 stage ring oscillator. The next section compares the circuit performance of TFETs with different Ge mole fraction and also compares them with corresponding similar dimension double gate MOSFET [18]. The threshold voltage of MOSFETs and TFETs are also kept the same for fair comparison.

## VI. Circuit Simulation Results

The models of tunnel FETs were used in simulation of simple inverter, nand gate, nor gate and 5 stage ring oscillator. The performance of circuits was evaluated on the grounds of dynamic power, leakage power and delay. All the circuit simulations are done at $V_{dd}$ = 0.5 V. The inverter, nand gate and nor gate were loaded with a 1pF load (much greater than internal capacitances).

*A. Delay*

A very general expression for dependence of delay on various parameters in a digital circuit is given by equation 3. Clearly delay depends on both total output capacitance and current. Since for the case of inveter, nand and nor gate we use a load capacitance which is much greater than internal capacitances (for both MOSFET and TFET). For instance load capacitance is 1pf compared to output internal capacitance of ~1 fF for a complete Ge TFET. Therefore total capacitance is constant for different devices and is nearly independent of the internal capacitance. Hence the delay is only dependent on current. However for a ring oscillator there is no external capacitance therefore the capacitance in delay expression changes for different devices hence delay is dependent on both capacitance (internal) and current.

$$Delay = K . \frac{Capac.}{Current} \qquad (3)$$

Fig. 16 compares the delay of TFETs for various Ge mole fraction in a loaded inverter , nand and nor gate. Clearly the delay decreases with increasing Ge mole fraction due to the increasing current. For an inverter delay is 0.014 μs for complete Ge TFET compared to 1.87 μs for a complete Si. Fig. 17 compares delay of TFET circuits with MOSFET circuits. MOSFETs are slightly advantageous over TFETs due to their lower delay as clear from Fig. 17. Delay of MOSFET invertercircuit for Ge=0.8 is 0.0058 μs compared to 0.014 μs for TFET inverter circuit for same Ge mole fraction. The scene changes when we analyze the 5 stage ring oscillator, Fig. 18. In this case the delay is dependent on both the internal capacitance and the current, since there is no load capacitance. For a 5 stage ring oscillator the TFETs outperform MOSFETs because in MOSFETs effect due to increasing current is compensated by the increasing capacitance hence their delay is more compared to TFET circuit. However among TFETs for different Ge mole fraction the delay of 5 stage ring oscillator decreases with increasing Ge mole fraction because the increase in current is more compared to the increase in capacitance.

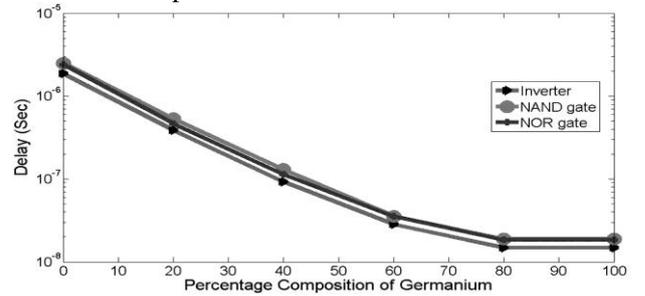

Fig. 16 Delay vs Ge composition for TFET. $V_{dd}$ = 0.5 V.

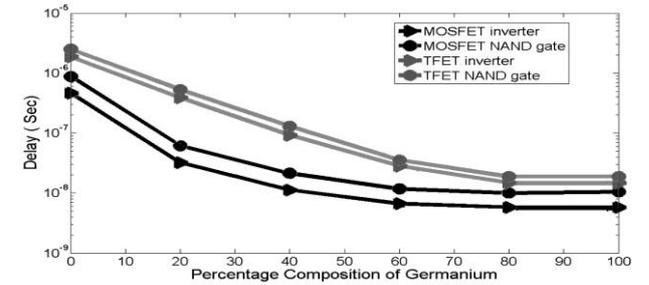

Fig. 17 Delay vs Ge composition. $V_{dd}$ = 0.5 V

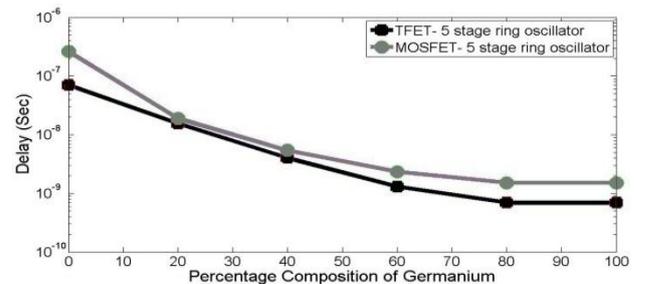

Fig. 18 Delay vs Ge composition. $V_{dd}$ = 0.5 V

### B. Dynamic power

Dynamic power is the power dissipated during the switching from on state to off state and *vice versa*. Fig.19 compares the dynamic power dissipated for a loaded inverter, nand and nor gate. Clearly dynamic power increases with increasing Ge mole fraction. Maximum dynamic power is given by equation 4. Since maximum frequency is limited by delay. And delay is inversely proportional to current. Finally we arrive at equation 5 which is independent of capacitance. Hence the dynamic power dissipated increases with Ge mole fraction due to increase in current with Ge mole fraction. For instance dynamic power for inverter with Ge=0 is $6.6 \times 10^{-9}$ W

$$power = k.freq_{min}.C.V_{dd}^2 \qquad (4)$$
$$= k_1.freq_{min}.C = k_2.C/delay$$
$$= k_2.\frac{current}{C}.C = k_2.current \qquad (5)$$

compared to $8.4 \times 10^{-7}$ W for pure Ge inverter. Fig. 20 compares the dynamic power of MOSFET circuits with

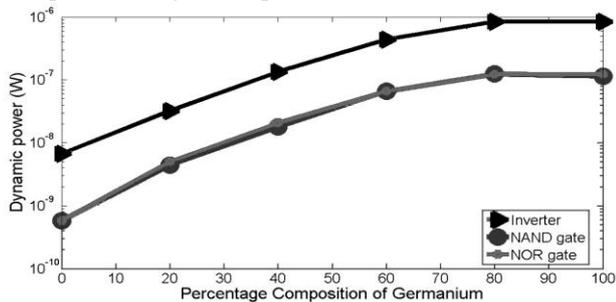
Fig. 19 Dynamic power vs Ge composition for TFET. $V_{dd} = 0.5$ V

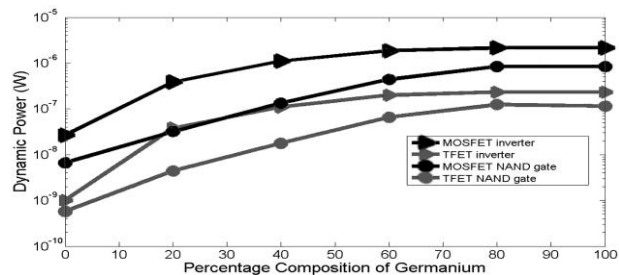
Fig. 20 Dynamic power vs Ge composition. $V_{dd} = 0.5$ V

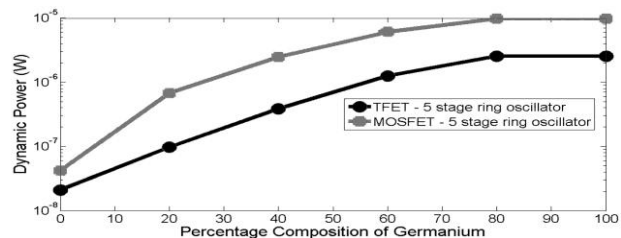
Fig. 21 Dynamic power vs Ge composition. $V_{dd} = 0.5$ V

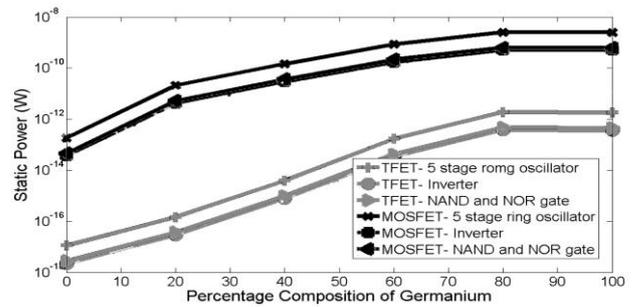
Fig. 22 Static power vs Ge composition. $V_{dd} = 0.5$ V

TFET circuits. Clearly dynamic power dissipated by MOSFETs circuits is more than TFET circuits due to their higher current. For Ge=0.8, MOSFET inverter circuit has dynamic power of $2.14 \times 10^{-6}$ W compared to $8.4 \times 10^{-7}$ W for corresponding TFET. Finally Fig. 21 compares the dynamic power dissipated in 5 stage ring oscillator for both MOSFET and TFET. For a ring oscillator also the dynamic power is more for MOSFET circuits since dynamic power is only dependent on current.

### A. Leakage Power

Leakage power or static power is the power consumed by the circuit when it is not switching, i.e., when it is in steady state. Static power is mainly constituted by the off current or leakage current of the devices in circuits. We compare the static power dissipated by TFET circuits with that of MOSFET circuits in Fig. 22. Clearly TFETs are highly advantageous over MOSFETs when it comes to static power dissipation due to their very low off current compared to MOSFETs. However this off current in TFETs increases with increasing Ge mole fraction due to lower band gap of SiGe with high Ge content and hence the leakage power also increases with Ge mole fraction.

## VII. Conclusion

In this paper we saw the advantage of extended channel SiGe TFET over normal double gate TFET. We were able to decrease the off current by gate under lap and thus were able to take advantage of the lower band gap of Ge to increase the on current. We also studied various characteristics of this TFET and compared it with similar dimension MOSFETs. We then evaluated the circuit performance of these TFETs for various Ge mole fractions and also compared them with corresponding MOSFETs. We saw that TFETs outperform MOSFETs in circuits with no or very low load. However for a high load MOSFETs are advantageous in terms of delay. TFETs outscore MOSFETs in terms of leakage power in any circuit. Performance of TFETs can be improved further by using lower band gap materials and by finding ways to decrease the off current.

## References

[1] K.Boucart and A.M. Ionescu, Length scaling of double gate tunnel fet with high K gate dielectric, *Solid-State Electronic* 51(2007) 1500-1507

[2] K.Boucart ,A.M. Ionescu, Double gate tunnel FET with ultra thin silicon body and high K dielectric,. *ESSDERC 2006:383-386*

[3] K.Boucart ,A.M. Ionescu, Double gate tunnel FET high K gate dielectric, *IEEE Trans Electron Dev 2007;54:1725-33*

[4] P.F. Wang et al., Complementary Tunneling Transistor for low power application, *Solid-State Electronics 48(2004) 2281-2286*

[5] N. Arakawa, Y.Otaka, K.Shikki, Evaluation of barrier height and thickness in tunneling junctions by numerical calculation on tunnel probability, *Thin Solid Films 505 (2006) 67-70*

[6] Bhuwalka K, Born M, Sedlmaier S, Schulzee J, Eisele I, P-channel tunnel field-effect transistors down to sub-50 nm channel lengths, *Japanese Journal of Applied Physics 2006;45;3106-9*

[7] Mathias Born et.al, Tunnel FET: a cmos device for high temperature applications, *Proc. 25th international conference on microelectronics, 14-17may, 2006.*

[8] P. Zhao, J. Chauhan and J.Guo, Computational study of tunneling transistor based on grapheme nanoribbon, *NANO Letters 2009. Vol.9, No.2, 684-688[i]*

[9] N. Patel , A. Ramesha, S.Mahapatra, Drive current boosting of n-type tunnel fet with strained SiGe layer at source, *Microelectronics Journal 39(2008) 1671-1677*

[10] Bhuwalka K, Schulzee J, Eisele I, Performance enhancement of vertical tunnel field effect transistor with SiGe in the p+ layer, *Japanese Journal of Applied Physics,Vol.43, No. 7A, 2004, pp.4073-4078*

[11] D. Kim, Y.Lee, Low power circuit design using based on hetro junction tunneling transistors (HETTs), *ISLPED'09 Aug,2009 San Francisco, California, USA*

[12] Hasanali G. Virani and Anil Kottantharayil, Optimization of Hetero Junction n-channel Tunnel FET with High-k Spacers, *2009 2nd international workshop on electron devices and semiconductor technology*

[13] F.Mayer, C. Le Royer et.al, Impact of SOI, $Si_{1-x}Ge_xOI$ and GeOI substrates on CMOS compatible Tunnel FET performance, *Electron Device Meeting 2008. IEDM 2008. IEEE International. 15-17 Dec. 2008.*

[14] *Taurus Medici, Medici User Guide,* Version Y-2006.06, June 2006

[15] Sneh Saurabh and M. Jagadesh Kumar, Impact of Strain on Drain Current and Threshold Voltage of Nanoscale Double Gate Tunnel Field Effect Transistor:Theoretical Investigation and Analysis, *Japanese Journal of Applied Physics 48(2009), Vol.48, paper no. 064503*

[16] Ioffe Physico-Technical Institute. *Electronic archive, New Semiconductor Materials. Characteristics and Properties (http://www.ioffe.ru/SVA/NSM/Semicond/SiGe/bandstr.html)*

[17] *OrCad PSpice,A/D User's Guide*, (Part number 60-30-636)

[18] Saurabh Mookerjea et.al, On Enhanced Miller Capacitance Effect in Internal Tunnel Transistors, *IEEE Electron Device Letters, Vol.30, No. 10. October 2009, 1102-04.*

## Acknowledgments

We would like to thank Ministry of Human Resource Development (MHRD), India, for sponsoring our research. A brief version of this paper has been accepted in "International conference on nanotechnology research and commercialization" (ICONT 2011), Sabah, Malaysia. This paper contains more than 50 percent additional material compared to the conference paper.